\documentclass[aps,prl,reprint,groupedaddress]{revtex4-1}
\usepackage{graphicx} 
\usepackage{dcolumn} 
\usepackage{bm}
\usepackage{hyperref}
\usepackage{siunitx}
\usepackage{float}
\usepackage{graphicx} 
\usepackage{epstopdf}
\usepackage{verbatim}
\usepackage{array}
\usepackage{booktabs}
\usepackage{multirow}
\usepackage{amsmath}
\usepackage{amssymb}

\begin{document}
\newenvironment{figurehere}
{\def\@captype{figure}}
{}
\makeatother

\title{Light fragment and neutron emission in high-energy proton induced spallation reactions}
\author{Hui-Gan Cheng}
\author{Zhao-Qing Feng}
\email{Corresponding author: fengzhq@scut.edu.cn}

\affiliation{School of Physics and Optoelectronics, South China University of Technology, Guangzhou 510640, China}

\date{\today}

\begin{abstract}
The dynamics of high-energy proton-induced spallation reactions on target nuclides of $^{136}$Xe, $^{59}$Ni, $^{56}$Fe, $^{208}$Pb, $^{184}$W, $^{181}$Ta, $^{197}$Au and $^{112}$Cd, are investigated with the quantum molecular dynamics transport model. The production mechanism of light nuclides and fission fragments is thoroughly analyzed. The statistical code GEMINI is employed in conjunction to the model for managing the decay of primary fragments. For the treatment of cluster emission during the preequilibrium stage, a surface coalescence model is implemented into the model. It is found that the available data of total cross sections are well reproduced with the combined approach for the spallation reactions on both the heavy and light targets, i.e., $^{56}$Fe and $^{208}$Pb, while it is underestimated in the intermediate-mass-fragment region for the medium-mass target $^{136}$Xe.
The energetic clusters are mainly contributed from the preequilibrium recognition, in which the quantum tunneling is taken into account. On the other hand, a fairly well overall description of light cluster and neutron emission is obtained and detailed discrepancies with respect to the experimental results are discussed. Possible modifications on the description of spallation reactions are stressed and compared with both recent experimental and theoretical results in the literature.

\begin{description}
\item[PACS number(s)]
25.40.Sc, 24.10.Lx, 24.10.Pa             
\end{description}
\end{abstract}

\maketitle

\section{I. INTRODUCTION}\label{intro}

Recently, both experiments and theories have witnessed their huge growth in the spallation reactions ever since the spallation neutron source proved itself a powerful tool in researches and applications \cite{source1, source2}. In the social and ecological areas, spallation reaction lies at the heart of the transmutation of long-lived radiotoxic nuclear waste whose half-life can be drastically shortened to an acceptable scale of hundreds of years via fast fission induced by neutrons. This process at the same time generates enough energy to supply the electric grid and sustain the facility itself \cite{ads1, ads2, ads3, ads4, ads5, ads6, ads7}. In space missions, preflight assessments on the damaging effects on the astronauts and electronic parts must be undertaken with spallation models \cite{space} in order for the success of a space flight. In astrophysics, spallation cross-section is a key input to the evaluation of the propagation of cosmic rays both in the atmosphere and inter-stellar media \cite{cosmicrays1, cosmicrays2, cosmicrays3}. In material science, neutrons produced by spallation neutron sources are used to probe the properties of condensed matter \cite{material1, material2}. Other areas include synthesis of rare isotopes \cite{synthesis1, synthesis2, synthesis3}, cancer therapy \cite{cancer1, cancer2, cancer3}, biology \cite{biology}, cosmography \cite{cosmography}, etc.. Because of the diversity of the applications as mentioned above, the broad range of reaction conditions including the beam energy and the target size, and the complicated reaction mechanisms that the spallation reaction tails, great challenges have been imposed on the experimental measurements of the yields and kinematics of spallation products (see \cite{latest} for latest advances), on the calculation of the related observables, and on the theoretical understanding of the underlying mechanisms through which the spallation products emerge. For a comprehensive review on these subjects, we refer the readers to Refs. \cite{overview1, overview2}.

Typically, the spallation reaction is described by the two-step model, as was first suggested by Seber \cite{serber}. In view of this description, in the first stage, the incident light particle at energies from hundreds of MeV to several GeV induces a cascade of collisions through a series of hadron-hadron collisions within a target way heavier than the incident projectile. This process has a momentary duration of only tens of fm/c, in which the incident energy is partly emitted in form of high energy ejectiles, pions, nucleons and Light-Charged-Particles with $Z \le 4$ (LCPs) for instance, while the remaining part is deposited during the process thermalizing the target nucleus, the latter often excited to hundreds of MeV. The first stage is a fast dynamical process, whereas the second stage is a statistical one and several orders of magnitude longer in duration than the former. During the second stage, random fluctuations in the distribution of energy and nuclear density multiply locally or globally, respectively leading the compound nucleus to undergo light particle evaporation or fission sequentially, until all products are fully deexcited. In the light of these considerations, very practical numerical codes based on the ideas of  intranuclear cascade plus statistical decay have already been developed \cite{incl1} and improved \cite{incl2} which are capable of reproducing yields and kinematics of some reaction products to an appreciable accuracy for spallation reactions at incident energies not lower than 200 MeV. However, on the theoretical ground as pointed out and discussed in Refs. \ \cite{imf2004,imf2007,imf2008}, more sophisticated reaction mechanism beyond the simple two-step model is required in order to account for the production of the Intermediate-Mass-Fragments (IMFs) featured in its experimentally revealed triple-humped kinematics in the velocity distributions \cite{triple hump}. So these features were studied by introducing the deexcitation mode of multifragmentation through Intranuclear-Cascade plus Statistical-Multifragmentation model (INC+SMM) \cite{imf2008} or more inherent models such as the Boltzmann-Langevin-One-Body (BLOB) \cite{blob} or the Quantum-Molecular-Dynamics model (QMD) \cite{fan zhang}. Moreover, the fermionic properties of nucleons also has an important role to play in the formation of IMFs \cite{imfpsdc } and thus the beginning part of discussion of this work is related to this area of research.

Following this, comes the central focus of this work which has a lot to do with application. In the design of shielding of spallation facilities and astronautical equipments, the energy spectra of LCPs at all angles in spallation reactions are a vital information \cite{space} and when experimental data are not available, theoretical simulations become indispensible. In microscopic transport models for nuclear reactions, one approach to treat the pre-equilibrium emission of LCPs and give a reliable account of the Double-Differential-Cross-sections (DDXs) at any angles is a modification of the transport model through incorporating a coalescence mechanism near the target surface in the whole of time evolution. This kinematical treatment was first discussed by Goldberger \cite{Goldberger} and Metropolis \cite{Metropolis} and later on modified by Nagle \cite{Nagle} and Mattiello {\it et al.} \cite{Mattiello}. Nowadays this mechanism has been mounted onto INCL \cite{INCL_surf1, INCL_surf2, INCL_surf3, INCL_surf4}, QMD \cite{QMD1_surf, QMD2_surf} and the coalescence exciton model \cite{EXI_surf}, for the refinement of LCPs production in these models. Thus in the second part of this work, dynamics of pre-equilibrium emission of light clusters are studied and discussed in the light of this idea and the results of calculations of the DDXs of LCPs are presented for LCPs with Z up to 2 and A up to 4.

This paper is organized as follows: Section II is a brief description of the models employed in this work. Section III is the presentation of the results of our calculation in which subsection 1 is devoted to the discussion on the reproduction of total yields of spallation fragments under various conditions and the results are compared with previous studies. In subsections 2 and 3, the DDXs of LCPs and neutrons are presented and discussed, followed by a brief summary in section IV.

\section{II. MODEL DESCRIPTION}

\subsection{1. Transport model}
The quantum molecular dynamics model was first proposed by J. Aichelin et al. \cite{aichelin1986} as a novel approach based on the idea of classical molecular dynamics model to incorporate the wave-particle duality of microscopic systems. Later on, the fermionic nature of nucleons was considered by solving the equation of motion starting right from the anti-symmetrized wave equation of the nucleus as a whole, i.e., the Fermionic-Molecular-Dynamics model (FMD) \cite{fmd1990 } or the Anti-symmetrized-Molecular-Dynamics model (AMD) \cite{amd1992}.

In this work, the Lanzhou-Quantum-Molecular-Dynamics (LQMD) code \cite{LQMD1, LQMD2, LQMD3} is employed, in which the motion of the individual nucleons is parameterized as Gaussian wave packets in both coordinate space and momentum space
\begin{eqnarray}
\phi_{i}(\mathbf{r}, t)=&&
\frac{1}{(2\pi \sigma_{r}^{2})^{3/4}} {\rm exp} [-\frac{(\mathbf{r}-\mathbf{r}_{i}(t))^{2}}{4\sigma_{r}^{2}}]
\nonumber\\
&& \cdot {\rm exp}(\frac{i\mathbf{p}_{i}(t)\cdot\mathbf{r}}{\hbar})
\end{eqnarray}
where $ \mathbf{r}_{i}(t)$ and $ \mathbf{p}_{i}(t)$ are the centers of the wave packets in coordinate space and momentum space separately. The width of the packets depends on the parameter $\sigma_{r}$. These are parameters to be solved by subjecting the following total wave function of the reaction system to the variational method \cite{variation}
\begin{eqnarray}
\Phi(\mathbf{r},t)=\prod_{i}\phi_{i}(\mathbf{r},\mathbf{r}_{i},\mathbf{p}_{i},t).
\end{eqnarray}
Neglecting the change of the packet width $L$ through time and letting them to be constants, the equations of motion of the wavepacket parameters $\mathbf{r}_{i}^{,} \ s$ and $\mathbf{p}_{i}^{,} \ s$ are obtained formally as
\begin{align}
\dot{\mathbf{r}_{i}}=\frac{\partial H}{\partial \mathbf{p}_{i}}, \quad
\dot{\mathbf{p}_{i}}=-\frac{\partial H}{\partial \mathbf{r}_{i}}
\end{align}
together with the density-functional Hamiltonian
\begin{eqnarray}
H=T+U_{Coul}+\int V_{loc}[\rho(\mathbf{r})] d \mathbf{r} +U_{MDI}
\end{eqnarray}
where $U_{Coul}$ is the Coulomb energy of the whole system and $V_{loc}$ is the nuclear potential energy density which is evaluated through Wigner transformation \cite{wigner}, and takes the form
\begin{eqnarray}
V_{loc}(\rho)=&& \frac{\alpha}{2}\frac{\rho^{2}}{\rho_{0}} +
\frac{\beta}{1+\gamma}\frac{\rho^{1+\gamma}}{\rho_{0}^{\gamma}} + E_{sym}^{loc}(\rho)\rho\delta^{2}
\nonumber \\
&& + \frac{g_{sur}}{2\rho_{0}}(\nabla\rho)^{2} + \frac{g_{sur}^{iso}}{2\rho_{0}}[\nabla(\rho_{n}-\rho_{p})]^{2}
\end{eqnarray}
where
\begin{align}
\rho(\mathbf{r},t) &= \int f(\mathbf{r}, \mathbf{p}, t)d\mathbf{p}    \nonumber \\
&=\sum_{i} \frac{1}{(2\pi \sigma_{r}^{2})^{3/2}} {\rm exp} \left[-\frac{(\mathbf{r} - \mathbf{r}_{i}(t))^{2}}{2\sigma_{r}^{2}}\right]
\\
\label{wigner-transform}
f(\mathbf{r},\mathbf{p},t)&=\sum_{i} f_{i}(\mathbf{r},\mathbf{p},t)   \nonumber \\
&=\sum_{i} \frac{1}{(\pi \hbar)^{3}} {\rm exp} \left[-\frac{(\mathbf{r} - \mathbf{r}_{i}(t))^{2}}{2\sigma_{r}^{2}} - \frac{(\mathbf{p} - \mathbf{p}_{i}(t))^{2}\cdot 2\sigma_{r}^{2}}{\hbar^{2}} \right],
\end{align}
while $U_{MDI}$ is the momentum dependent interaction (MDI) \cite{mdi} and assumes the form
\begin{eqnarray}
U_{MDI}=&& \frac{1}{2\rho_{0}}\sum_{i,j,j\neq i}\sum_{\tau,\tau'}C_{\tau,\tau'}\delta_{\tau,\tau_{i}}\delta_{\tau',\tau_j}\int\int\int d\textbf{p}d\textbf{p}'d\textbf{r} \nonumber \\
&& \times f_{i}(\textbf{r},\textbf{p},t)\left[\texttt{ln}(\epsilon(\textbf{p}-\textbf{p}')^{2}+1)\right]^{2}f_{j}(\textbf{r},\textbf{p}',t)
\end{eqnarray}
respectively.

The coefficients of each term are the mean-field parameters constrained by reproducing the basic saturation properties and the incompressibility within a sensible range for symmetric nuclear matter. Two sets of mean-field parameters labelled PAR1 and PAR2 are chosen for the calculations, as given in Table \ref{skyrme} along with their incompressibilities. In the MDI term, $C_{\tau, \tau^{'}}=C_{mom}(1+x)$ for $\tau=\tau^{'}$ and $C_{\tau, \tau^{'}}=C_{mom}(1-x)$ for $\tau\neq\tau^{'}$, where the subscripts $\tau$ and $\tau^{'}$ stand for isospin whose values are -1 and 1, and the parameter $x=-0.65$ is the strength of isospin splitting. In the isospin asymmetric terms,  $\rho_{n}$, $\rho_{p}$ and $\rho=\rho_{n}+\rho_{p}$ are the neutron, proton and total densities, respectively, $\delta=(\rho_{n}-\rho_{p})/(\rho_{n}+\rho_{p})$ being the isospin asymmetry. The coefficients in the isospin-dependent and density-gradient-dependent terms $g_{sur}$, $g_{sur}^{iso}$ and $\rho_{0}$ are taken to be 23 MeV fm$^{2}$, -2.7 MeV fm$^{2}$ and 0.16 fm$^{-3}$, respectively.

In addition to the motion under the nucleons' mean field, the collision between nucleons is another key ingredient that makes up the time evolution of the reaction system. In the simulation, when the spacial separation of any two nucleons in their center-of-mass frame is smaller than a value
\begin{eqnarray}
r_{NN} = \sqrt{\sigma_{NN}(\sqrt{s})/\pi},
\end{eqnarray}
a collision between the two nucleons is considered, in which $\sigma_{nn}(\sqrt{s})$ is the total nucleon-nucleon collision cross-section at their invariant mass $\sqrt{s}$. The NN elastic scattering cross-section is parameterized to fit the experimentally available data in a wide energy domain \cite{Fe09}. Finally, taking into account the effect of Pauli-blocking due to the fermionic property of nucleons, the collision is decided to be executed or blocked by comparing with a random number the blocking probability $b_{ij}=1-(1-\overline{P}_{i})(1-\overline{P}_{j})$ of the two participant nucleons $i$ and $j$ in the final state in which $\overline{P}_{i}$ is given by
\begin{eqnarray}
\overline{P}_{i}=  && \frac{32\pi^{2}}{9h^{3}}\sum_{i\neq k, \tau_{i}=\tau_{k}} (\Delta r_{ik})^{2}(3R_{0}-\Delta r_{ik})  \nonumber \\
&&  \times(\Delta p_{ik})^{2}(3P_{0}-\Delta p_{ik}).
\end{eqnarray}
The $\Delta r_{ik}=|\mathbf{r}_{i}-\mathbf{r}_{k}|$ and $\Delta p_{ik}=|\mathbf{r}_{i}-\mathbf{r}_{k}|$ are the relative distances of two nucleons in coordinate and momentum spaces, respectively. The summation is satisfied the criterion in phase space $\Delta r_{ik}<2R_{0}$ and $\Delta p_{ik}<2P_{0}$ with $R_{0}=3.367$fm and $P_{0}=112.5$ MeV/c.

\begin{table*}[htp]
	\vspace{20pt}
	\centering
	\caption{Skyrme parameters PAR1 and PAR2 employed in the LQMD model.}
    \setlength{\tabcolsep}{1.5em}
	\begin{tabular}{cccccccc} \hline
		\specialrule{0em}{1pt}{1pt}
		\centering
		 & $\alpha \ ({\rm MeV})$  & $\beta \ ({\rm MeV})$ & $\gamma$ & $C_{mom} \ ({\rm MeV})$ & $\epsilon \ ({\rm c^{2}/MeV^{2}})$ & $m^{*}_{\infty}/m$ & $K_{\infty} \ ({\rm MeV})$ \\
		\hline
				\specialrule{0em}{1pt}{1pt}
		PAR1 & -215.7 & 142.4 & 1.322 & 1.76 & $5 \times 10^{-4}$ & 0.75 & 230 \\
				\specialrule{0em}{1pt}{1pt}
		PAR2 & -226.5 & 173.7  & 1.309 & 0. &0. & 1. & 230 \\
		\hline
		
		\label{skyrme}
	\end{tabular}
\end{table*}

\subsection{2. Fragment recognition and statistical decay}
At the end of dynamical evolution when all violent changes have settled and the nucleons are re-aggregated and condensed to form individual clusters, a procedure called Minimum Spanning Tree (MST) must be followed to identify these hot remnants before the transition to statistical decay. In the LQMD model, a constituent nucleon can incorporate into its intermediate cluster a neighboring nucleon of relative momentum and location $\Delta p \leq 200$ MeV/c and $\Delta r \leq 3.5$ fm with respect to itself, given that this new nucleon is also located close around the surface of the cluster, i.e., with a distance smaller than 3.5 fm plus the r.m.s radius of the cluster. Also two neighboring intermediate clusters can join to form a bigger cluster if the size of the new cluster they thus compose is within a limit which is adopted as the liquid-drop-model radius.

After the hot remnants are reconstructed, the simulation is conveyed to the next stage, cooling down by statistical decay. The statistical decay is realized by the GEMINI code by R. J. Charity \cite{charity}. Generally speaking, in the GEMINI code, the compound nucleus experiences a sequence of binary divisions in forms of light particle evaporation or fission, until the compound nucleus is thoroughly deexcited. In asymmetric divisions, as for the emission of light particles with $Z$ up to 4, the Hauser-Feshbach formulism is adopted \cite{hauser} and the decay width of emitting a light particle $(Z_{1},A_{1})$ of spin $J_{1}$ from a mother nucleus $(Z_{0},A_{0})$ of excitation energy $E^{*}$ and $J_{0}$ leaving behind it a residue $(Z_{2},A_{2})$ of spin $J_{2}$ is given by,
\begin{align}
	\Gamma_{J_{2}}(Z_{1},A_{1},&Z_{2},A_{2})= \nonumber \\
	\displaystyle{\frac{2J_{1}+1}{2\pi \rho_{0}}} & \sum_{l=|J_{0}-J_{2}|}^{J_{0}+J_{2}} \int_{0}^{E^{*}-B-E_{rot}(J_{2})} T_{l}(\epsilon)\rho_{2}d\epsilon
\end{align}
where $\rho_{0}$ and $\rho_{2}$ are the level densities of the mother and the residual nucleus, respectively, and $T_{l}(\epsilon)$ is the transmission coefficient. $B$ is the binding energy between the light particle and the residue and $E_{rot}(J_{2})$ the rotation plus deformation energy of the latter. For asymmetric fission, Moretto's generalized transition-state formalism \cite{moretto} which determines the fission probability by the phase space density on the ridge line around the saddle point is used. For symmetric fission which is an available option in the code's input, the Bohr-Wheeler formalism \cite{bohr} is used. Fission barrier heights are mainly calculated through the rotating-finite-range model \cite{rfr} and both shell and pairing corrections are also considered.

\subsection{3. Phase space density constraint method}
Nucleons are fermions and so nucleons with the same isospin are repulsive to one another in avoidance of getting to close in phase space. In the QMD framework, this effect was simply neglected or counted in by introducing an artificial phase space repulsive potential \cite{pauli1, pauli2} until M. Papa proposed \cite{papa2001} that this may be solved by performing unphysical elastic collisions for nucleons coming too close to each other in phase space, maintaining in a local phase space occupation at the center of each nucleon $i$, $\overline{f}_{i} < 1$ with $\overline{f}_{i}$ in the form
\begin{eqnarray}
\overline{f}_{i}=\sum_{j}\delta_{s_{i},s_{j}}\delta_{\tau_{i},\tau_{j}}\int_{h^{3}}f_{j}(\mathbf{r},\mathbf{p},t)d\mathbf{r}d\mathbf{p}
\end{eqnarray}
where $f_{j}(\mathbf{r},\mathbf{p},t)$ as already implied in Eq. \ref{wigner-transform} is the Wigner transform of the nucleon $j$'s wavepacket and the subscript $s$'s and $\tau$'s respectively stand for the spin and isospin of the corresponding nucleons. The above described method is called Phase-Space-Density-Constraint (PSDC). Since the Pauli exclusion effect is repulsive between identical fermions, with the PSDC included, the expansion and multifragmentation phenomenon in some reactions underestimated by the model are supposed to be better reproduced \cite{imfpsdc }.

\subsection{4. Surface coalescence}
For a better description of pre-equilibrium cluster emission, we alloyed the surface coalescence model into the LQMD model following the specifications given by Ref. \cite{INCL_surf1}. When an outgoing nucleon trepasses a certain radial distance $R_{0} + D_{0}$ with respect to the center of the mother nucleus, recursive construction of LCPs from this leading nucleon is initiated by picking up a first nucleon, a second and a third and so on. Here $R_{0} = 1.4A_{targ}^{1/3}$ fm and for the proton incident energies involved, $D_{0}$ is taken to be a proper value 2 fm. In our work, only the constructions of $d$, $t$, $^{3}$He and $^{4}$He clusters are considered. Of course, this method has already been extended to include the construction of heavier clusters \cite{NewPot}, which, for a preliminary reliability test, is not yet considered in the present work. During the process, an intermediate cluster picks up a nucleon to form higher clusters by judging the following phase space condition,
\begin{eqnarray}
R_{Nj} P_{Nj} \le h_{0}, \quad R_{Nj} \ge 1 fm
\end{eqnarray}
where $R_{Nj}$ is the spatial distance between the intermediate cluster $N$ and the nucleon $j$ to pick up ,and $P_{Nj}$ is the relative momentum between the two objects. Let $\mathbf{R}_{N}$ and $\mathbf{r}_{j}$ be the position of the intermediate cluster and the nucleon in coordinate space, $\mathbf{p}_{N}$ and $\mathbf{p}_{j}$ the momenta, and $M_{N}$ and $m_{j}$ the masses of the two objects. They have the form,
\begin{gather}
R_{Nj} = \displaystyle{|\mathbf{R}_{N} - \mathbf{r}_{j}|}  \nonumber\\
P_{Nj} = \displaystyle{|\frac{m_{j}}{M_{N} + m_{j}}\mathbf{p}_{N} - \frac{M_{N}}{M_{N} + m_{j}}\mathbf{p}_{j}|}.
\end{gather}
The latter is in fact the momentum of either object in their common center-of-mass frame. As to the choice of the phase space parameter $h_{0}$, though various refinements are available, for simplicity, we adopted those prescribed by Ref. \cite{INCL_surf1} and Ref. \cite{INCL_surf2} which we label by Set 1 and Set 2 as listed in Table \ref{Surf_coal_para}. When all possible combinations including $d$, $t$, $^{3}$He and $^{4}$He constituted by different nucleons and the leading nucleon are listed. An emission test is performed according to the priority $^{4}$He, $^{3}$He (or $t$),  $d$, in which a $^{4}$He particle is randomly selected among all other $^{4}$He possibilities and tested to see if its total energy under the target mean field renders its penetration through the Coulomb plus Woods-Saxon barrier. If the candidate passes the test, it is emitted along its tangential direction and the time evolution of the reaction system is resumed. Otherwise a lower cluster in the priority list is selected in the same way and tested and so on. If all the tests fail, the penetration test is performed on the leading nucleon to decide whether it is emitted or reflected. The total energy of all emission candidates are calculated according to the following equation,
\begin{eqnarray}
E_{lcp} = \sum_{i=1}^{A_{lcp}} (E_{i} + V_{i}) + B_{lcp}
\end{eqnarray}
where $E_{i}$ and $V_{i}$ are the kinetic energy and potential energy of the constituent nucleon $i$ under the target mean field, $B_{lcp}$ is the binding energy of the cluster, and $A_{lcp}$ is the mass number of the cluster. Last but not least, in the procedure stated above, all clusters constructed must be appropriately far away from the center of the target nucleus in order that they be clusters formed near the target's surface and $R_{l}$ measures this distance which is taken to be
\begin{eqnarray}
R_{l} = CA_{targ}^{1/3}.
\end{eqnarray}
A too small $C$ results in a too rich production of clusters and vice versa. About this, there is a brief discussion in the corresponding section.

\begin{table}[h]
	\vspace{20pt}
	\centering
	\caption{Suface coalescence parameters Set 1 and Set 2}
	\setlength{\tabcolsep}{2.5em}
	\begin{tabular}{lcc}		
		\hline
		\specialrule{0em}{1pt}{1pt}
		Construction & \multicolumn{2}{c}{$h_{0}$ (fm MeV/c)} \\ \cline{2-3}
		\hline
		\specialrule{0em}{1pt}{1pt}		
		& Set 1 & Set 2 \\
		\hline
		\specialrule{0em}{1pt}{1pt}
		$p + n \rightarrow d$				& 387 & 336 \\
		$d + n \rightarrow t$				& 387 & 315 \\
		$d + p \rightarrow ^{3}$He			& 387 & 315 \\
		$t + p \rightarrow ^{4}$He			& 387 & 300 \\
		$^{3}$He $+ n \rightarrow ^{4}$He	& 387 & 300 \\
		\hline
		
		\label{Surf_coal_para}
	\end{tabular}
\end{table}

\subsection{5. Simulation settings}
For any one reaction system in this calculation, the maximum impact parameter $b_{max}$ is chosen as $b_{max}^{'}+0.3$ fm where $b_{max}^{'}$ is the smallest impact parameter at which the target no longer suffers from nucleon-nucleon collision with the incident proton passing by. The extra 0.3 fm is reserved for Coulomb excitation. Beside the maximum impact parameter, the switching time from dynamical stage to statistical stage is another determining factor for a reliable reproduction of the realistic physical circumstances. In INC simulation, this quantity is given by an established formula \cite{incl1}, which in our case, however, is not proper for QMD simulation since the latter is capable of describing the oncoming evolution after the system has been fully excited. The criterion we adopted to select the switching time is such that after this moment of time, the all observables in question be relatively stable as time goes on after the end of the violent fluctuation of the preceding dynamical evolution. Furthermore, during the pre-equilibrium cascade process, nucleon and cluster emitted in the forward direction are generated in an earlier stage whereas those in the backward direction emerge in a later stage. Because of this, the pre-equilibrium time span must be long enough as to cover the emissions at all polar angles. Considering all these complications and for a shorter CPU time, the switching times of $p + ^{56}$Fe, $^{58}$Ni are 65 fm/c, $^{112}$Cd, $^{136}$Xe, 85 fm/c, and $^{181}$Ta, $^{184}$W, $^{208}$Pb, 115 fm/c.

\section{III. RESULTS AND DISCUSSION}
\subsection{1. Total yields of spallation fragments}
There has long been a debate on the physical origin of the sources of IMFs in spallation reactions. Fission and multifragmentation were proposed as source candidates hinted by experimental revelation \cite{triple hump}. When the incident energy is low or impact parameters are large, the amount of incident energy deposited in the target nucleus is small and the excited nucleus undergoes normal fission-evaporation deexcitation mode to cool down.

\begin{figure*}
	\includegraphics[width=16 cm]{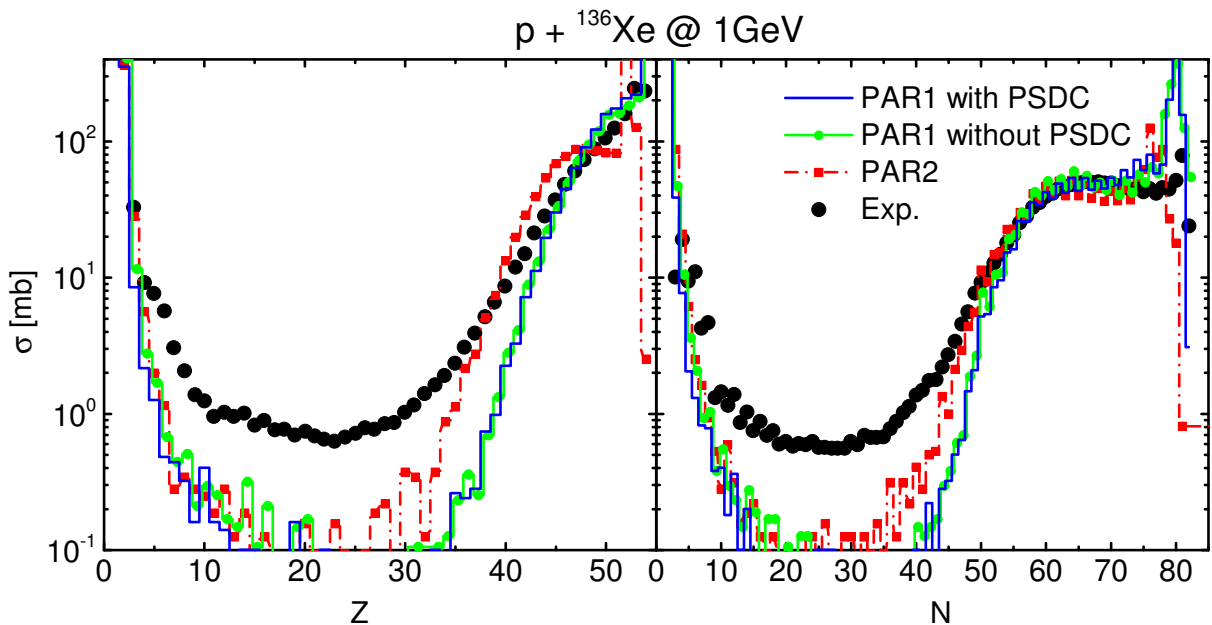}
	\vspace{-0.3cm}
	\caption{The fragment yields as functions of the charge number (left panel) and the neutron number (right panel) for in the reaction of $p+^{136}$Xe at the incident energy of 1 GeV. The experimental data are taken from Ref. \cite{data136Xe}. }
	\label{cnpcs1}
\end{figure*}

\begin{figure}
	\includegraphics[width=8 cm]{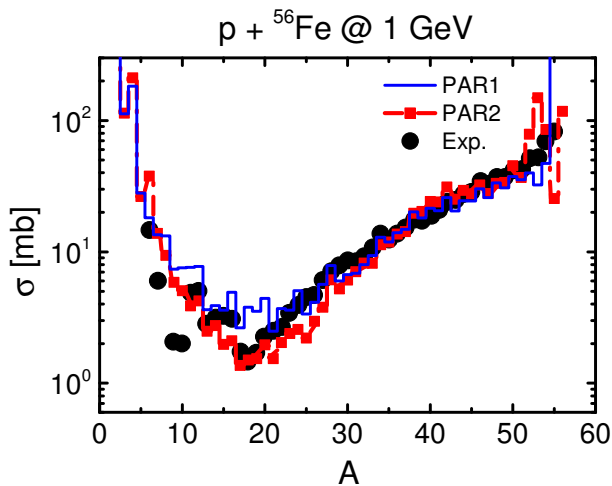}
	\vspace{-0.35cm}
	\caption{Calculated total cross-section as a function of mass number for $p+^{56}$Fe at 1 GeV. The experimental data are taken from Ref. \cite{data56Fe}.}
	\label{cnpcs2}
\end{figure}

However, when the excitation energy exceeds about a certain threshold \cite{threshold}, the hot nucleus may expand and a fast breakup becomes possible. These two different ways of deexcitation would exhibit different kinematics in their velocity spectra. In the velocity spectra of IMFs, contributions from both mechanism has been disentangled into separate kinematic components \cite{triple hump}, which provides a very strong evidence of the existence of multifragmentation in proton-induced spallation reactions. This section can partly be deemed as an investigation of the IMFs production, following some of the previous studies in the literature.

The total cross-section $\sigma$ of the reaction system $p + ^{136}$Xe at 1 GeV plotted as functions of the charge number Z and the mass number A is presented in Fig. \ref{cnpcs1}. The experimental data, taken from Ref. \ \cite{data136Xe}, are plotted in black solid dots. In this part, we tried to employ the PSDC method in the reproduction of the IMFs yields, as was previously done in Ref. \cite{fan zhang} in which the reproduced IMFs cross-sections in between $5\le Z\le40$ agrees with the experimental data to some extent. Our calculation, differing from that of Ref. \cite{fan zhang} in the mean field parameters (given in the first row of Table \ \ref{skyrme}) and the technical details in the code, turned out to be quite different. As expected, without PSDC (shown in blue histogram), the IMFs yields are seriously underestimated. In comparison to this, the introduction of the PSDC method only negligibly increases the IMF cross-sections (as shown by the green histogram in comparison to the blue one in Fig. \ref{cnpcs1}). This may suggest that the PSDC method does help filling the blank between $5\le Z \le40$, but the efficiency of the PSDC method in inducing multifragmentation of the excited target nucleus may depend greatly on the choice of the mean field parameters and the technical details in the code. On the other hand, it is found that when a set of mean field parameters PAR2 which incorporates the momentum dependent interaction is used instead, the production of target-like fragments are underestimated together with the spetra right below it overestimated. This can be scribed to the extra fluctuation brought in for the presence of the MDI which causes spurious emission of nucleons in the pre-equilibrium stage.

Now the investigation with the same purposes is extended to a lighter reaction system $p + ^{56}$Fe at 1 GeV, as shown in Fig. \ref{cnpcs2}. It is seen that the experimental results \cite{data56Fe} are well reproduced under either momentum dependent or independent mean fields in both the trend and value to some extent. This time, however, the IMFs yields turn out to be overestimated on an overall scale without MDI, at the cost of the underestimation of the target-like tail. Nevertheless, the local trends are same for both settings. The fact that the success of the reproduction of the IMFs under either condition may be ascribed to the relatively more abundant multifragmentation, evaporation or relatively more cascade emission allowed at higher per-nucleon excitation energies gained at the same incident proton energy but with atomic mass twice smaller compare to the previous case of $^{136}$Xe. The differences between the results of the two settings might be understood that in calculation with MDI, the IMFs formed during the reaction is more unstable out of more violent fluctuations and thus more nucleons but less IMFs are emitted with respect to the case without MDI. For analysis on the experimental data and the results of INC calculation of IMFs production with the same reaction system, see Ref. \cite{imf2008}.

\begin{figure*}
	\includegraphics[width=17 cm]{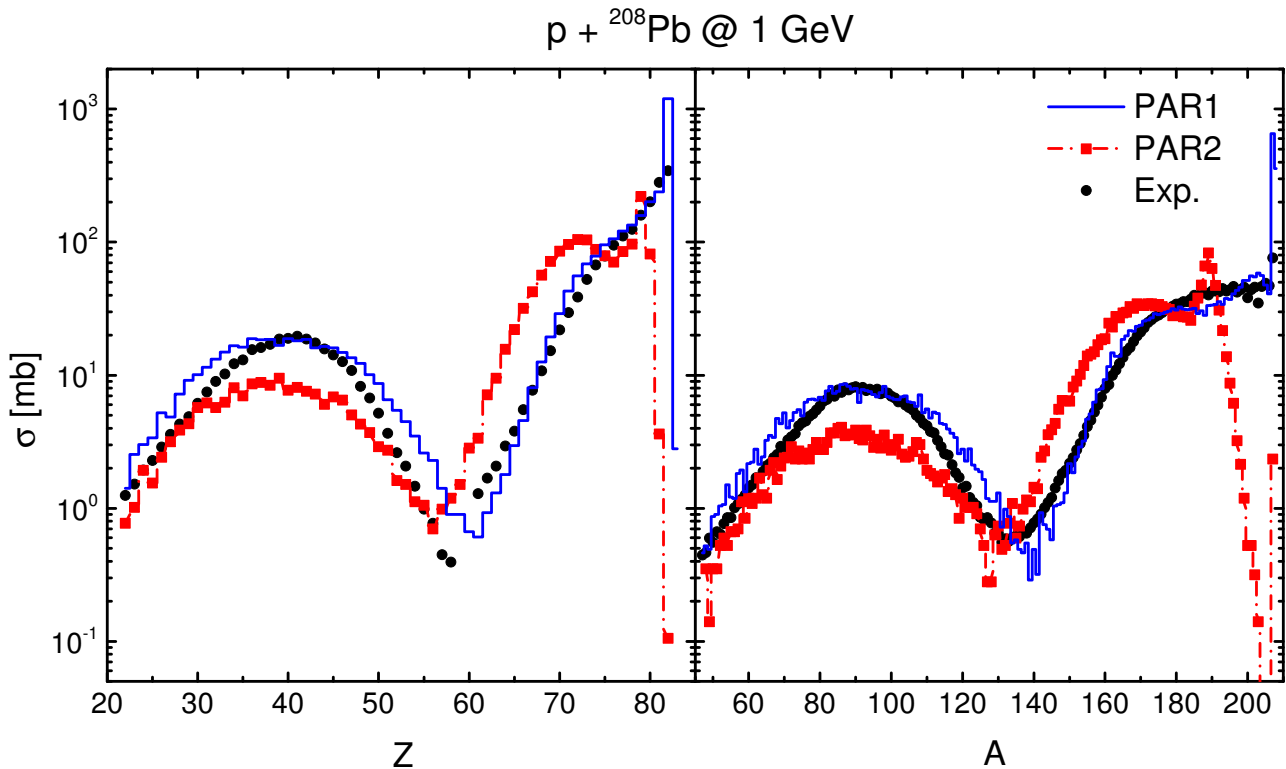}
	\vspace{-0.33cm}
	\caption{Calculated total cross-section plotted versus the charge number (left) and the mass number (right) for $p+^{208}$Pb at 1 GeV. The experimental data are taken from Ref. \cite{data208Pb}. Here the fission delay parameters prescribed in Ref. \cite{theo208Pb_OuLi} are adopted instead of the default ones.}
	\label{208Pb}
\end{figure*}

In the preceding discussion, $^{56}$Fe, a light target was considered. In the next, we draw our attention to another extreme, high-energy proton induced spallation reaction at still the same energy but on a very heavy target, $^{208}$Pb with which more exhibited divergences between simulations with and without momentum dependent interaction be expected as the fission process comes to play a key role. In Fig. \ref{208Pb}, the total cross-section plotted as a function of both the atomic number Z and the mass number A is presented. The black dots stand for the experimental data which are taken from Ref. \cite{data208Pb}, the red line for calculation with MDI and the blue line for the case without MDI. It is seen that above all, calculation with or without MDI both reproduce the main trend and the main features of the experimentally measured spectra but it is apparent that the case without MDI gives a very much better overall fit to the data whereas the case with MDI peaks too early at the target-like end in the plot versus the mass number, which is accompanied by an overestimation of the region between the target-like end and the valley in the middle of the graph. This is due to the spurious emission of nucleons in the presence of MDI which always tends to induce more fluctuation. As a result, the fission peak is underestimated since the very target-like residues which possess lower fission barriers and are thus more fission-likely ended up rarer with respect to the case without MDI. To end the present discussion, let us make one more final comment on the results of the case without MDI. In the statistical decay stage of our simulation, the fission delay parameters as previously prescribed in Ref. \cite{theo208Pb_OuLi} are adopted instead of the default ones in GEMINI. In Fig. \ref{208Pb}, it is seen that our results are roughly the same as those given by Ref. \cite{theo208Pb_OuLi}. The heights of both the fission peak and the target-like tail of the spectra agree very well with respect to the experimental data. So our results can serve as a further confirmation of the fission delay prescription given by Ref. \cite{theo208Pb_OuLi}.

\subsection{2. Light cluster kinetic energy spectra}
\begin{figure*}
	\includegraphics[width=16 cm]{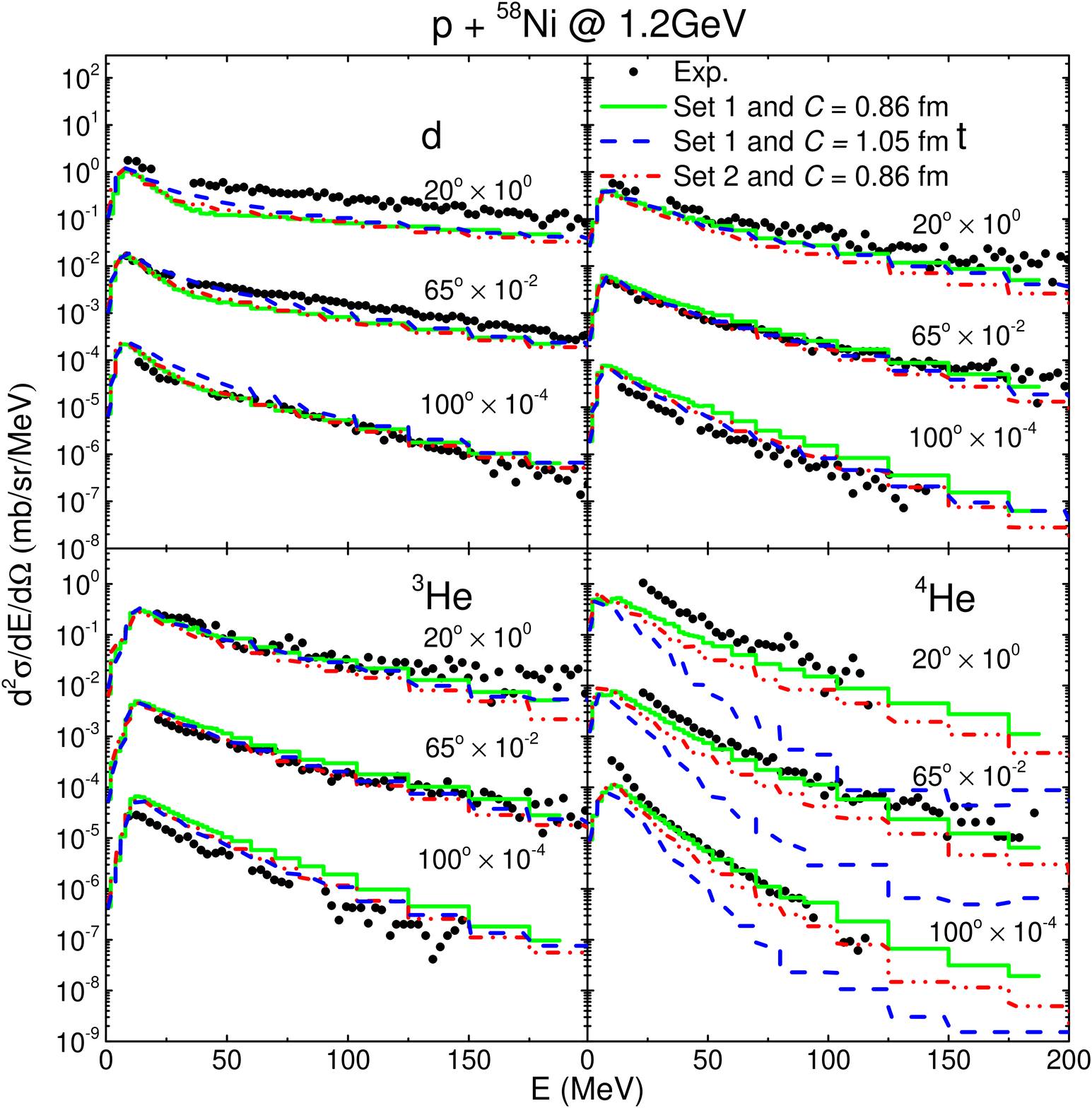}
	\vspace{-0.1cm}
	\caption{Double differential cross-section of $d$, $t$, $^{3}$He and $^{4}$He as a function of kinetic energy and polar angle for p + $^{58}$Ni at 1.2 GeV calculated with different sets of parameter. The data are taken from Ref. \cite{Parafit}.}
	\label{58Ni}
\end{figure*}

\begin{figure*}
	\resizebox{1\textwidth}{!}{
		\includegraphics{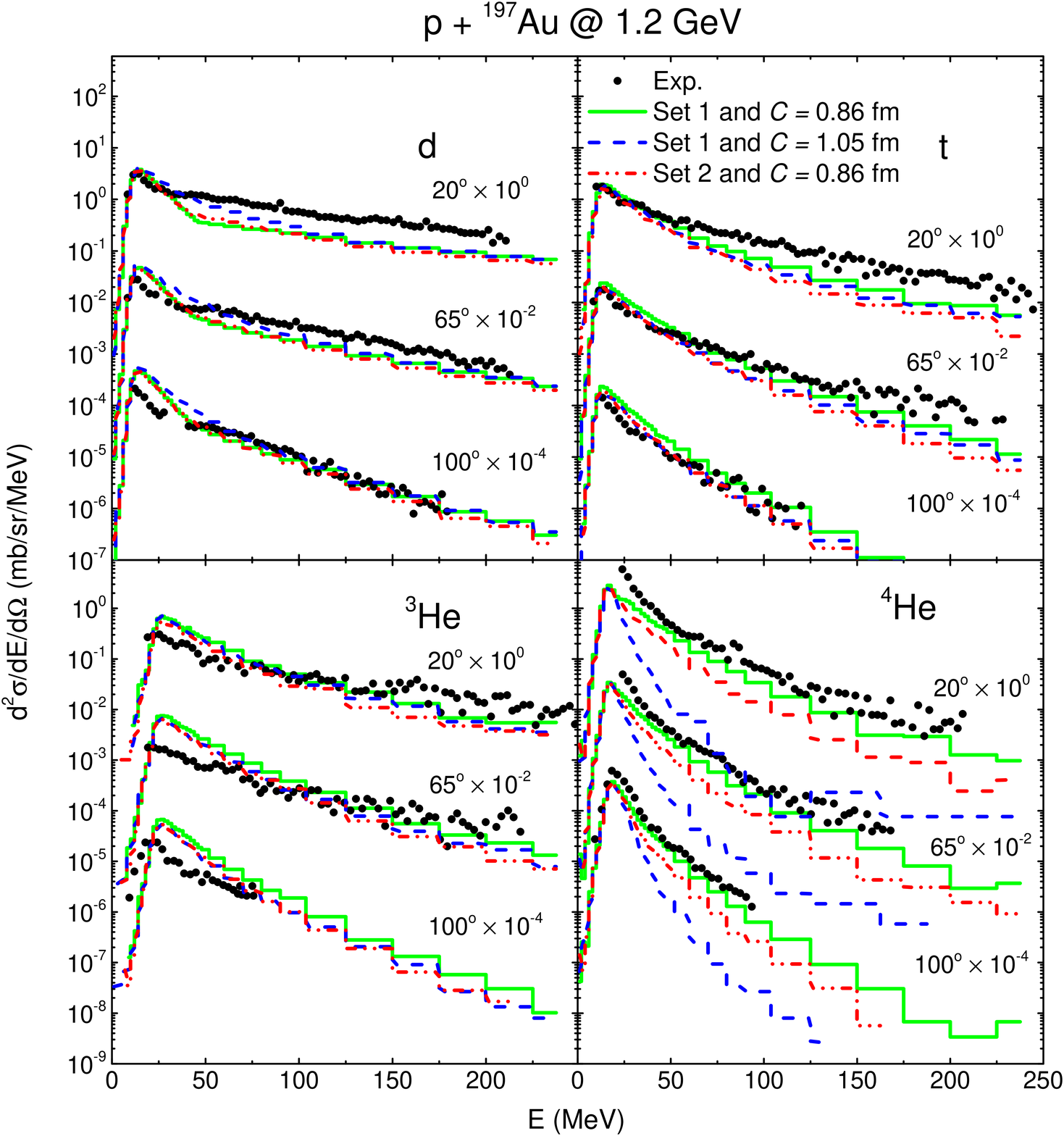}
	}
	\vspace{-0.6cm}
	\caption{Double differential cross-section of $d$, $t$, $^{3}$He and $^{4}$He as a function of kinetic energy and polar angle for p + $^{197}$Au at 1.2 GeV calculated with different sets of parameter. The data are taken from Ref. \cite{Parafit}.}
	\label{197Au}
\end{figure*}

\begin{figure*}
	\resizebox{1\textwidth}{!}{
		\includegraphics{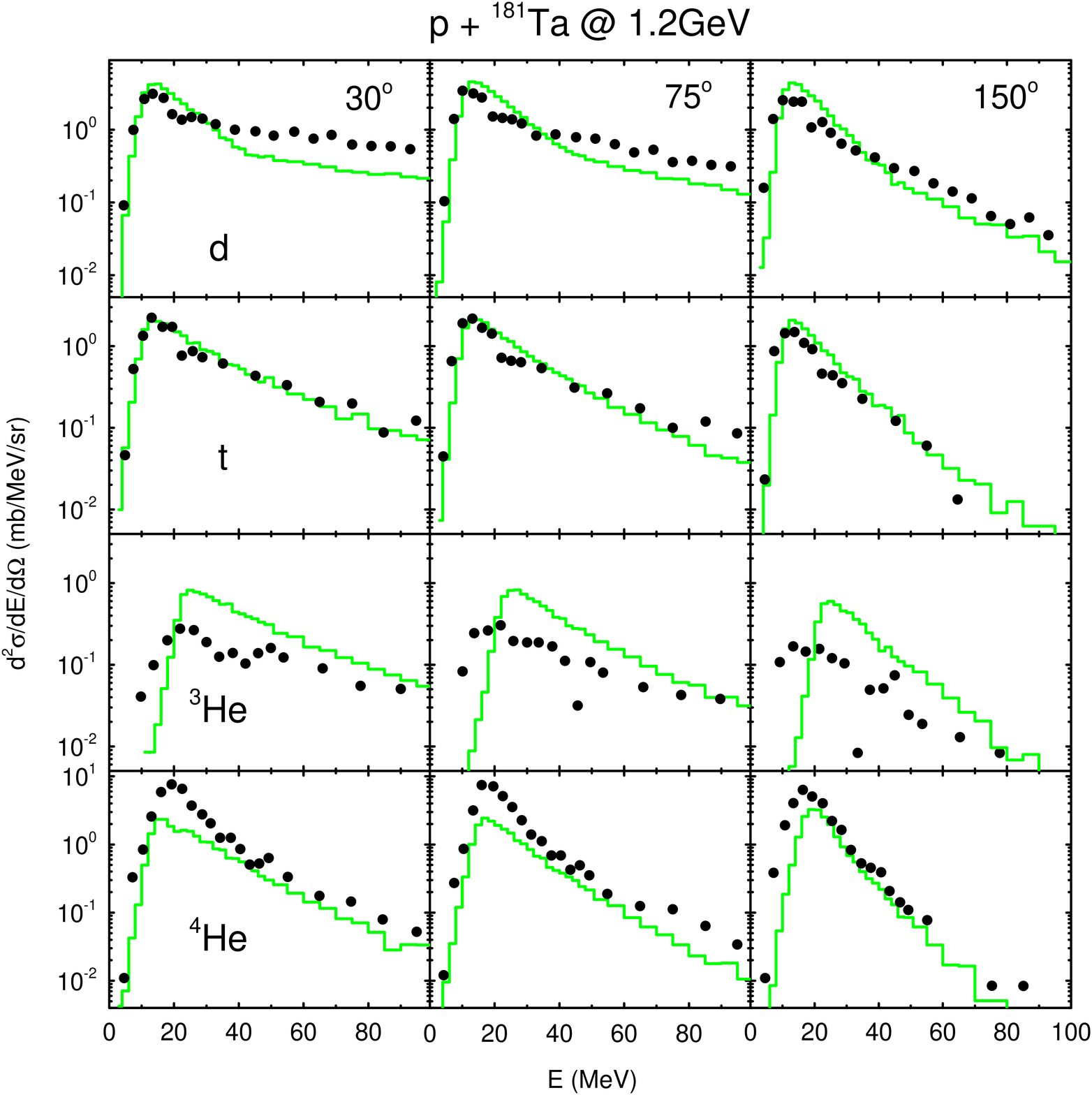}
	}
	\vspace{-0.55cm}
	\caption{Double differential cross-section of $d$, $t$, $^{3}$He and $^{4}$He as a function of kinetic energy and polar angle for p + $^{181}$Ta at 1.2 GeV calculated with $C = 0.86$ fm and phase space parameters Set 1. The data are taken from Ref. \cite{INCL_surf3}.}
	\label{181Ta}
\end{figure*}

For the production of high energy LCPs in high-energy proton induced spallation reaction, we considered the following three reaction systems: p + $^{58}$Ni at 1.2 GeV, p + $^{181}$Ta at 1.2 GeV and p + $^{197}$Au at 1.2 GeV whose experimental data are available. Since the present work is intended to provide an overall description of the spallation reaction and a test of the predictive power of the LQMD model in such reaction scenario, we did not dig deeper into the vast and arduous task of parameter fitting to the experimental results as was already done in a recent work \cite{Parafit}. So we just make a few comments on the results we so far obtained. In Fig. \ref{58Ni} and Fig. \ref{197Au}, the DDXs of light cluster production at three different angles are presented for targets $^{58}$Ni and $^{197}$Au bombarded on by proton at 1.2 GeV, the values being scaled by a 10$^{-2}$ for every angle with respected to the former one. Besides, similar results are presented in Fig. \ref{181Ta} for p + $^{181}$Ta at 1.2 GeV.

We see that in most cases, the high energy tails of the DDXs are reproduced very well except for the very forward angles of $d$ and for large angles of $t$ and $^{3}$He. So it turns out that with a rather rough set of parameters, a fairly acceptable quality of description can still be obtained for both light and heavy targets bombarded on by high-energy protons. However it is noticed that the region around the potential barrier is sometimes overestimated and other descrepancies with respect ot the experimental data are present. They can arise from the following sources. Firstly, the production of light clusters in the surface coalescence model is regulated by two types of parameters, the distance coefficient $C$ which controls the separation of constructed clusters from the center of the target nucleus, and the phase space parameter $h_{0}$ which controls the size of the clusters in phase space. In Fig. \ref{58Ni} and Fig. \ref{197Au}, the calculation with three different choices of parameters are plotted by lines in different colours and shapes as indicated by the legends. It can be observed that increasing the threshold separation of the constructed clusters from the target's center by increasing the parameter $C$ results in, to some extent, a similar effect as that of substituting the phase space parameters Set 2 for Set 1 which sets a larger upper bound for the phase space sizes of the constructed clusters. Both settings bring down the high energy tails, except for $d$. However, for an agreeable reproduction of the high energy tails of large clusters with respect to experiments , say $^{4}$He cluster for example, $C$ must not be too large. Otherwise the high energy tails of these clusters die out too early. Secondly, a high quality description of the emergence of the leading nucleons that initiate the construction of clusters is a prerequisit for a high quality description of cluster production and meanwhile, a correct time evolution of the phase space nuclear density distribution of the target nucleus is also important. As seen in the next section, our reproduction of the neutron DDXs is not that desirable quantitatively for backward angles, which can acount for the corresponding discrepancies that occur in our cluster DDXs. Thirdly, in our consideration of barrier tunneling, the contribution of centrifugal potential to the total barrier height is neglected. Some outgoing clusters with energy around the barrier sometimes bring away with them ten to twenty units of angular momentum measured in $\hbar$ and that amounts to a contribution of, for instance, for a $^{4}$He cluster with $l=10$ in a $^{197}$Au target, about 6 MeV to the barrier height and thus this modifies both the height and shape of the spectra around the barrier. Apart from the interplay among all these above, other potential sources may also be able to acount for the problems. Nevertheless, there are a lot more efforts required to give a higher quality reproduction of cluster DDXs \cite{HighQual}, but a simple surface coalescence model with these roughly selected parameters, implemented into our LQMD model, works rather well in describing the key characteristics of light cluster emission in spallation reactions.

\begin{figure*}
	\resizebox{1\textwidth}{!}{
		\includegraphics{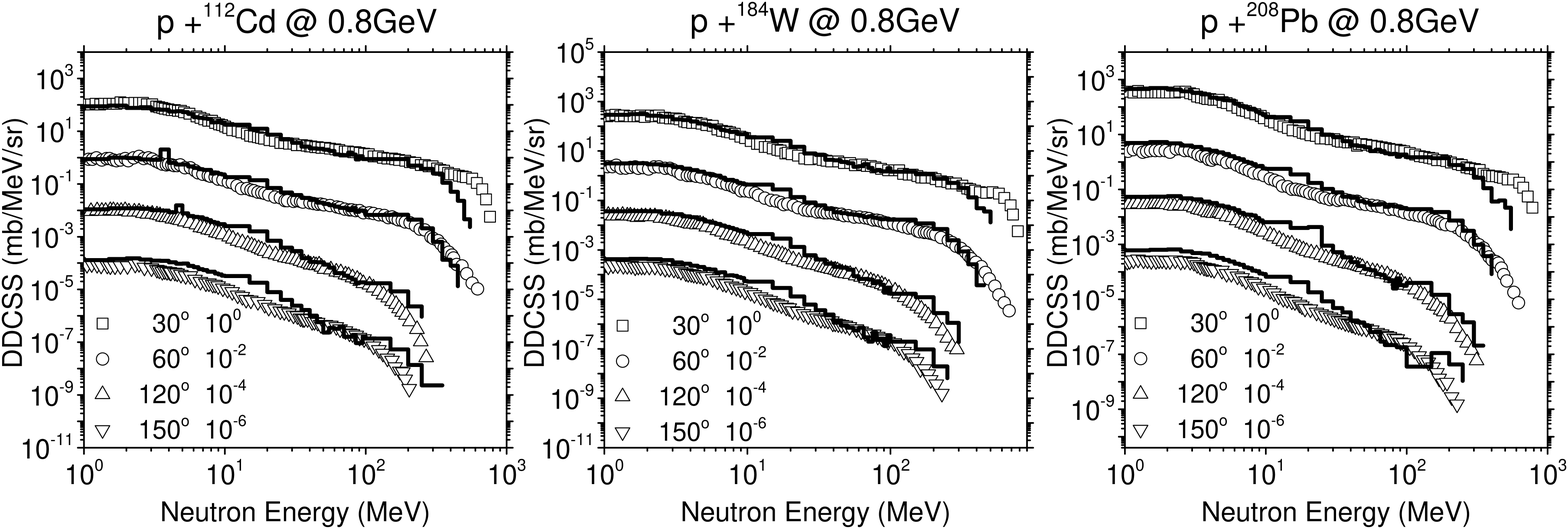}
	}
	\vspace{-0.6cm}
	\caption{Calculated neutron DDXs of $p+^{112}$Cd, $^{184}$W, $^{208}$Pb at 800 MeV. The experimental data are taken from Ref. \cite{DDXs data}.}
	\label{neutron}
\end{figure*}

\subsection{3. Neutron double differential cross-sections}
In this section, the model is applied to the reproduction of the DDXs of spallation neutron which has been intensely investigated experimentally and theoretically on a large number of spallation targets over a vast range of incident energies in the past decades. The accurate neutron DDXs is a vital information for the design and the various utilizations of the spallation neutron source \cite{application of DDXs }. However, whenever experimental data are not available, theoretical calculation tools, such as the moving source model \cite{moving source }, Intranuclear-Cascade plus Evaporation model (INC+E) \cite{inc+e} or HETC-3STEP \cite{hetc-3step} play indispensable roles. QMD calculation of the DDXs, as a more sophisticated way than any other, was studied by G. Peilert {\it et al.} \cite{QMDspallation1} who were soon followed by K. Niita {\it et al.} \cite{QMDspallation2,QMDspallation2,QMDspallation3,QMDspallation4}. Comparison of the results of calculation among different sets of mean field parameters has already been studied by Li Ou {\it et al.} in Ref. \cite{Li Ou}.

In this section, the mean field parameters PAR1 are employed to simulate 800 MeV proton-induced spallation reactions on $^{112}$Cd, $^{184}$W and $^{208}$Pb targets. It is obvious in Fig. \ref{neutron} that the model can reproduce the main trend of the spectra given by experiments and the data are taken from Ref. \cite{DDXs data}. However, in the low energy domain around $E=20$ MeV, the data are somewhat overestimated with the increase of the polar angle, which we find is not originated from the evaporation stage but simply from the cascade stage in form of extra free neutrons. The high energy tails close to the incident energy at \ang{30} drop too soon, while at larger angles, it is rather nicely reproduced. The ambiguity of the results at energies close to the incident energy is simply due to the quality of statistics which is limited by the computational resources available.

\section{IV. SUMMARY}
Several aspects of high-energy proton induced spallation reactions have been investigated with the LQMD model, i.e., the total fragment yields and the DDXs of light clusters and neutrons, for different targets. For the total yields, it is found that the efficiency of the PSDC method in enhancing multifragmentation are dependent on the choice of mean field parameters and the technical details in the code. The PSDC method is favorable for multifragmentation and thus contribute to the total yields of IMFs. On the other hand, it is found that the inclusion of the MDI distorts the yield spectra of spallation on $^{136}$Xe and $^{208}$Pb. The agreement with the experimental data is obtained in terms of total fragment yields with the set of fission delay parameters in the GEMINI code, which again fortifies the validity of this prescription in this scenario. For the description of cluster emission from statistical decay and pre-equilibrium stages, the GEMINI code and a simple surface coalescence model are employed. Though the parameters adopted in the surface coalescence model are rough, a rather good overall reproduction of the DDXs of light clusters is achieved with our model and it is seen that there is a good potentiality that the model could be refined by polishing the choice of the different parameters and the potential barrier. For the reproduction of neutron DDXs, three heavy targets $^{112}$Cd, $^{184}$W, $^{208}$Pb and an incident energy of 800 MeV are chosen. The model reproduces the experimental results, but descrepancies remain to some extent, which is a push for future furtherance of the model in this aspect.

\section{Acknowledgements}

This work is supported by the National Natural Science Foundation of China (Projects No. 11722546 and No. 11675226) and the Talent Program of South China University of Technology.

\end{document}